# High-$T_C$ Superconductivity in Ultra-thin Crystals: Implications for Microscopic Theory

Dale R. Harshman[1,2,3*] and Anthony T. Fiory[4]

[1]*Physikon Research Corporation, Lynden, WA 98264, USA;*
[2]*Department of Physics, University of Notre Dame, Notre Dame, IN 46556, USA;*
[3]*Department of Physics, Arizona State University, Tempe, AZ 85287, USA;*
[4]*Department of Physics, New Jersey Institute of Technology, Newark, NJ 07102, USA*

**Abstract**

High transition temperature (high-$T_C$) superconductivity is associated with layered crystal structures. This work considers superconductivity in ultra-thin crystals (of thickness equal to the transverse structural periodicity distance *d* for one formula unit) of thirty-two cuprate, ruthenate, rutheno-cuprate, iron pnictide, organic, and transuranic compounds, wherein intrinsic optimal (highest) transition temperatures $T_{C0}$ (10 − 150 K) are assumed. Sheet transition temperatures $T_{CS} = \alpha T_{C0}$, where $\alpha < 1$, are determined from Kosterlitz-Thouless (KT) theory of phase transitions in two-dimensional superconductors. Calculation of $\alpha$ involves superconducting sheet carrier densities $N_S$ derived theoretically from crystal structure, ionic valences, and known doping, a two-fluid model for the temperature dependence of the superconducting magnetic penetration depth, and experimental data on KT transitions; $\alpha$ is on average 0.83 (varying with standard deviation 0.11). Experiments on several thin crystal structures of thickness $d_F$ approaching *d* are shown to be consistent with calculations of $T_{C0}$ from microscopic superconductivity theory and with $T_{CS}$ from KT theory, where the presence of disorder is also taken into account; careful analyses of these thin film studies indicate a minimum thickness $d_F \approx d$ for superconductivity.

## 1. Introduction

Differing significantly from conventional superconducting metals (Bardeen *et al.*, 1957), the high transition temperature (high-$T_C$) superconductors (Bednorz and Müller, 1986) are noted for a superconducting condensate with two-dimensional (2D) character, the absence of case II coherence factor effects, e.g. no Hebel-Slichter anomaly in nuclear magnetic resonance (Walstedt *et al.*, 1987, 1991; Warren *et al.*, 1987; Chubkov *et al.*, 2008; Parker *at al.*, 2008), a vanishingly small isotope effect (as one approaches optimization) (Harshman *et al.*, 2008, 2009) and other very unique features (Harshman and Mills, 1992). The two-dimensionality is evidenced by large anisotropies in the normal and superconducting electronic transport, and non-metallic transport in the direction perpendicular to the layered crystal structure. Superconducting properties can be optimized via doping or applied pressure to yield highest transition temperature and bulk Meissner fraction. Other materials properties that prove to be important from technological as well as scientific perspectives are relatively poor malleability and tendencies towards fracture, because ionic forces dominate crystal bonding (see Su *et al.* (2004) and references therein). Since their discovery (Bednorz and Müller, 1986), the family of high-$T_C$ superconductors has grown to include the cuprates [e.g. $YBa_2Cu_3O_{6.92}$ (Kamal *et al.*, 1998), $Bi_2Sr_2CaCu_2O_8$ (Sunshine *et al.*, 1988) and $La_{2-x}Sr_xCuO_4$ (Radaelli *et al.*, 1994)], rutheno-cuprates [e.g. $RuSr_2GdCu_2O_8$ (Jurelo *et al.*, 2007)], ruthenates such as $Ba_2YRu_{1-x}Cu_xO_6$ (Parkinson *et al.*, 2003), certain organic superconducting compounds [e.g. κ–[BEDT-TTF]$_2$Cu[NCS]$_2$ or κ–[BEDT-TTF]$_2$Cu[N(CN)$_2$]Br (Kini *et al.*, 1990)], various iron pnictide (and related) superconductors [e.g. $La(O_{1-x}F_x)FeAs$ (de La Cruz *et al.*, 2008)], and possibly transuranics [e.g. $PuCoGa_5$ (Wastin *et al.*, 2003)]. Optimal transition temperatures, distinguished by the notation $T_{C0}$, span a range from ~10 K to ~150 K.

Suppositions of pairing mechanisms based upon lattice vibrations in the high-$T_C$ superconductors have led to serious contradictions with experiment (Bourne *et al.*, 1987; Hoen *et al.*, 1989; Radousky, 1992; Gurvitch and Fiory, 1987; Gurvitch *et al.*, 1988; Harshman *et al.*, 2008, 2009). Consequently, we have focused our investigations toward Coulombic (electronic) pairing mechanisms; this possibility was suggested earlier by the observed systematic correlation between $T_{C0}$ and 2D carrier concentration $n_{2D}$, which is unique to optimally doped high-$T_C$ compounds (Harshman and Mills, 1992). A particularly relevant structural trait, pointing to a Coulombic origin of the superconductivity, and common to high-$T_C$ superconductors is the presence of at least two different types of charge layers; this is found, for example, in the cuprates, where the superconducting state can be created by oxygen doping or by substituting ions of different valences in different layers. A well known example of cation doping is $Sr^{+2}$ substitution for $La^{+3}$ in $La_{2-x}Sr_xCuO_{4-\delta}$; an example of anion doping is $Ca^{+2}$ substitution for $Y^{+3}$ in $(Y_{1-x}Ca_x)Ba_2Cu_3O_{7-\delta}$ (Böttger *et al.*, 1996); and an example of both cation and anion doping is exhibited by the compound $(Pb_{0.5}Cu_{0.5})Sr_2(Y_{0.6}Ca_{0.4})Cu_2O_{7-\delta}$ (Tang *et al.*, 1991).

These two types of charge layers comprise two types of charge reservoirs, which we denote as types I and II; for $YBa_2Cu_3O_{7-\delta}$ the type I charge reservoirs are composed of the BaO-CuO-BaO layers and the type II charge reservoirs contain the $CuO_2$-Y-$CuO_2$ structures. Given the recent research which unambiguously shows that mobile electrons coexist with the superconducting holes in $YBa_2Cu_3O_{7-\delta}$ (Harshman *et al.*, 2011a), and combining this with conclusions made in Harshman and Mills (1992), it is reasonable to assign these opposing charges to the type I and type II charge reservoirs, where (at least in the case of the p-type compounds) the pairing of the holes is mediated by the electrons across the interaction distance $\zeta$ (Harshman *et al.*, 2011b). This high-$T_C$ structure, which repeats with periodicity $d$ in the direction transverse to the layers and corresponds to one formula unit, is represented in Figure 1 for two example structures. The upper panel typifies compounds with trilayer reservoirs, where the interacting charge layers (solid circles) and doping ions (open circles) are grouped by charge reservoir type (colour). The lower panel typifies compounds with bilayer type I and monolayer type II reservoirs. The distance $\zeta$ separates the nearest-neighbor ions (of the outer layers in multiple-layer reservoirs) of adjacent type I and type II reservoirs. Assuming structures like those in Figure 1 and following simple charge sharing and scaling rules (relative to $YBa_2Cu_3O_{6.92}$, optimal stoichiometry), to determine $n_{2D}$ for each of 31 high-$T_C$ compounds, it was found that $k_B T_{C0} = \beta (N_{int})^{1/2} \zeta^{-1}$, where $\beta = 10.75 \pm 0.03$ eV nm$^2$ is universally constant among all the high-$T_C$ compounds and $N_{int}$ is the 2D superconducting interaction density (Harshman *et al.*, 2011b). The experimentally observed $T_{C0}$ and theoretically calculated $T_{C0}^{Th}$ values are presented in Table 1.

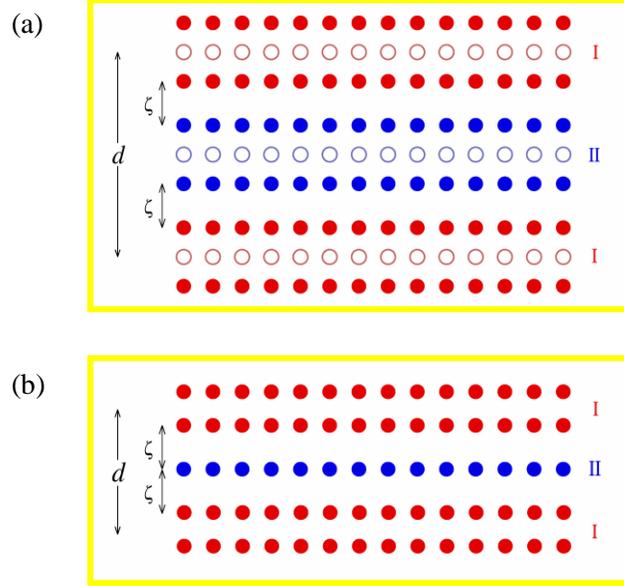

Figure 1. Stylized model structures of high-$T_C$ superconductors representing $YBa_2Cu_3O_{7-\delta}$ (upper) and $La_{2-x}Sr_xCuO_4$ (lower) after Harshman *et al.* (2011b). Cross section view perpendicular to basal plane of periodic electronic layers of types I (red circles) and II (blue circles). The distance $\zeta$ is the separation between adjacent nearest-neighbor layers of opposite type and $d$ is the periodicity.

For the high-$T_C$ pairing mechanism developed by Harshman *et al.* (2011b), the mediating and superconducting carriers are both contained within the unit cell; actually the type I and type II reservoirs are contained with the periodicity $d$. While the pairing interaction is associated with the adjacent mediating/superconducting bilayers, the other components of the charge reservoirs provide the necessary charge transport and equilibration mechanisms. Thus, absent any long-range 3D carrier mobility requirements, superconductivity should persist for layer thicknesses $d_F \geq d$. Experiments performed on $Bi_2Sr_2(Ca,Y)Cu_2O_8$ superlattices (Kanai *et al.*, 1990), ultrathin $Bi_2Sr_2CaCu_2O_8$ films (Sugimoto *et al.*, 1991; Saito and Kaise, 1998), one-unit cell $YBa_2Cu_3O_{7-\delta}$ (Terashima *et al.*, 1991; Matsuda *et al.*, 1992; Cieplak *et al.*, 1994) and $La_{2-x}Sr_xCuO_4$ multilayer films grown by molecular beam epitaxy (MBE) with layer-selective Sr and Zn doping (Logvenov *et al.*, 2009) have validated the 2D character of the superconducting condensate and provide good evidence for superconductivity in layers nearing unit-cell thickness. Results from several of these experiments are discussed in section 4.

Other indicators of intrinsic two-dimensionality in high-$T_C$ superconductors are found in the evidence for superconducting Kosterlitz-Thouless (KT) transitions (Beasley *et al.*, 1979; Halperin and Nelson, 1979). Previously examined in superconducting thin films of disordered $AlO_x$ (Hebard and Fiory, 1980), $InO_x$ (Fiory *et al.*, 1983), Hg-Xe (Kadin *et al.*, 1983) and crystalline $YBa_2Cu_3O_{7-\delta}$ (Fiory *et al.*, 1988), KT transitions within the individual layers occur at a temperature $T_{KT} < T_{C0}$ for weakly interacting layers, as shown by Clem (1991). Signatures of KT transitions have been found in transport experiments on $Bi_2Sr_2CaCu_2O_8$ crystals (Martin *et al.*, 1989), *c*-axis oriented thin films of $Tl_2Ba_2CaCu_2O_8$ (Kim *et al.*, 1989), La$_{2-x}$Sr$_x$CuO$_4$ (Kitano *et al.*, 2006), and unit-cell YBa$_2$Cu$_3$O$_{7-\delta}$ clad by Pr-doped YBa$_2$Cu$_3$O$_{7-\delta}$ (Matsuda *et al.*, 1992).

In section 2 we provide a brief theoretical background of Kosterlitz-Thouless theory, along with the necessary equations and assumptions used in our application to the high-T$_C$ systems. The relevant transition temperatures, screening lengths, etc., are calculated, tabulated and plotted for 32 high-T$_C$ compounds in section 3. Reported experimental data for testing the theory are analyzed in section 4, results are discussed in section 5, and we provide our conclusions in section 6.

**Table 1.** High-T$_C$ superconductor compounds listed in descending order of optimal transition temperature T$_{C0}$, with the corresponding interlayer periodicity distance *d* (after Harshman et al., 2011b). Calculated quantities are theoretical $T_{C0}^{Th}$, sheet charge density N$_{S0}$, sheet superconducting screening length $\Lambda_{S0}$, sheet transition temperature T$_{CS}$, product $\Lambda_{S0}$T$_{C0}$ and ratio $\alpha$ = T$_{CS}$/T$_{C0}$. Numbering sequence corresponds to abscissa of Figure 2.

| No. | Superconducting Compound | T$_{C0}$ (K) | $T_{C0}^{Th}$ (K) | *d* (nm) | N$_{S0}$ (nm$^{-2}$) | $\Lambda_{S0}$ (µm) | $\Lambda_{S0}$T$_{C0}$ (cm-K) | $\alpha$ (T$_{CS}$/T$_{C0}$) | T$_{CS}$ (K) |
|---|---|---|---|---|---|---|---|---|---|
| 1 | HgBa$_2$Ca$_2$Cu$_3$O$_{8+\delta}$ (25 GPa) | 145 | 144.5 | 1.43582 | 3.342 | 25.9 | 0.3753 | 0.9019 | 130.8 |
| 2 | HgBa$_2$Ca$_2$Cu$_3$O$_{8+\delta}$ | 135 | 134.3 | 1.57782 | 3.080 | 28.1 | 0.3792 | 0.9009 | 121.6 |
| 3 | TlBa$_2$Ca$_2$Cu$_3$O$_{9+\delta}$ | 133.5 | 132.1 | 1.5871 | 3.088 | 28.0 | 0.3740 | 0.9023 | 120.5 |
| 4 | Tl$_2$Ba$_2$Ca$_2$Cu$_3$O$_{10}$ | 130 | 130.3 | 1.794 | 3.076 | 28.1 | 0.3656 | 0.9046 | 117.6 |
| 5 | HgBa$_2$CaCu$_2$O$_{6.22}$ | 127 | 125.8 | 1.223 | 4.231 | 20.4 | 0.2596 | 0.9330 | 118.5 |
| 6 | (Bi,Pb)$_2$Sr$_2$Ca$_2$Cu$_3$O$_{10+\delta}$ | 112 | 113.0 | 1.8541 | 1.558 | 55.5 | 0.6218 | 0.8332 | 93.3 |
| 7 | Tl$_2$Ba$_2$CaCu$_2$O$_8$ | 110 | 108.5 | 1.4659 | 3.068 | 28.2 | 0.3101 | 0.9196 | 101.2 |
| 8 | YBa$_2$Cu$_4$O$_8$ (12 GPa) | 104 | 103.2 | 1.29042 | 3.210 | 26.9 | 0.2803 | 0.9275 | 96.5 |
| 9 | TlBa$_2$CaCu$_2$O$_{7-\delta}$ | 103 | 104.9 | 1.2754 | 3.066 | 28.2 | 0.2906 | 0.9248 | 95.3 |
| 10 | LaBa$_2$Cu$_3$O$_{7-\delta}$ | 97 | 98.0 | 1.1818 | 2.974 | 29.1 | 0.2821 | 0.9271 | 89.9 |
| 11 | HgBa$_2$CuO$_{4.15}$ | 95 | 92.2 | 0.95073 | 4.030 | 21.5 | 0.2039 | 0.9477 | 90.0 |
| 12 | YBa$_2$Cu$_3$O$_{6.92}$ | 93.7 | 96.4 | 1.16802 | 3.069 | 28.2 | 0.2641 | 0.9318 | 87.3 |
| 13 | Bi$_2$Sr$_2$CaCu$_2$O$_{8+\delta}$ (unannealed) | 89 | 86.6 | 1.5445 | 1.555 | 55.6 | 0.4952 | 0.8688 | 77.3 |
| 14 | Tl$_2$Ba$_2$CuO$_6$ | 80 | 79.9 | 1.16195 | 3.051 | 28.4 | 0.2268 | 0.9417 | 75.3 |
| 15 | Pb$_2$Sr$_2$(Y,Ca)Cu$_3$O$_8$ | 75 | 76.7 | 1.57334 | 1.557 | 55.6 | 0.4167 | 0.8906 | 66.8 |
| 16 | (Pb$_{0.5}$Cu$_{0.5}$)Sr$_2$(Y,Ca)Cu$_2$O$_{7-\delta}$ | 67 | 67.7 | 1.1829 | 1.173 | 73.7 | 0.4940 | 0.8692 | 58.2 |
| 17 | YBa$_2$Cu$_3$O$_{6.60}$ | 63 | 64.8 | 1.17279 | 1.344 | 64.4 | 0.4056 | 0.8936 | 56.3 |
| 18 | La$_{1.8}$Sr$_{0.2}$CaCu$_2$O$_{6\pm\delta}$ | 58 | 58.4 | 0.96218 | 0.696 | 124.4 | 0.7212 | 0.8048 | 46.7 |
| 19 | Sm(O$_{0.65-y}$F$_{0.35}$)FeAs | 55 | 56.3 | 0.42328 | 0.566 | 152.8 | 0.8402 | 0.7705 | 42.4 |
| 20 | (Sm$_{0.7}$Th$_{0.3}$)OFeAs | 51.5 | 51.9 | 0.42164 | 0.484 | 178.6 | 0.9200 | 0.7475 | 38.5 |
| 21 | RuSr$_2$GdCu$_2$O$_8$ | 50 | 50.3 | 1.15652 | 0.774 | 111.8 | 0.5591 | 0.8509 | 42.5 |
| 22 | Tb(O$_{0.80-y}$F$_{0.2}$)FeAs | 45 | 45.7 | 0.4166 | 0.336 | 257.8 | 1.1599 | 0.6790 | 30.6 |
| 23 | (Sr$_{0.9}$La$_{0.1}$)CuO$_2$ | 43 | 41.4 | 0.34102 | 0.320 | 270.0 | 1.1609 | 0.6787 | 29.2 |
| 24 | La$_{1.837}$Sr$_{0.163}$CuO$_{4-\delta}$ | 38 | 37.5 | 0.66029 | 0.574 | 150.8 | 0.5731 | 0.8470 | 32.2 |
| 25 | (Ba$_{0.6}$K$_{0.4}$)Fe$_2$As$_2$ | 37 | 36.9 | 0.66061 | 0.654 | 132.2 | 0.4890 | 0.8706 | 32.2 |
| 26 | Ce(O$_{0.84-y}$F$_{0.16}$)FeAs | 35 | 37.2 | 0.43016 | 0.252 | 343.4 | 1.2018 | 0.6673 | 23.4 |
| 27 | Ba$_2$YRu$_{0.9}$Cu$_{0.1}$O$_6$ | 35 | 32.2 | 0.41618 | 0.289 | 299.6 | 1.0488 | 0.7105 | 24.9 |
| 28 | Bi$_2$(Sr$_{1.6}$La$_{0.4}$)CuO$_{6+\delta}$ | 34 | 34.8 | 1.21995 | 0.345 | 250.8 | 0.8526 | 0.7669 | 26.1 |
| 29 | La(O$_{0.92-y}$F$_{0.08}$)FeAs | 26 | 24.8 | 0.43517 | 0.124 | 699.0 | 1.8174 | 0.5122 | 13.3 |
| 30 | Ba(Fe$_{1.84}$Co$_{0.16}$)As$_2$ | 22 | 23.5 | 0.64897 | 0.255 | 339.2 | 0.7462 | 0.7976 | 17.5 |
| 31 | PuCoGa$_5$ | 18.5 | 20.0 | 0.3397 | 0.112 | 769.3 | 1.4233 | 0.6072 | 11.2 |
| 32 | κ–[BEDT-TTF]$_2$Cu[N(CN)$_2$]Br | 10.5 | 11.6 | 1.47475 | 0.105 | 826.7 | 0.8680 | 0.7625 | 8.0 |

## 2. Two-dimensional superconductors

A superconducting material in the form of a thin sheet or film may be treated as a two-dimensional (2D) superconductor when its thickness $d_F$ is small compared to the superconducting magnetic penetration depth $\lambda_{\parallel}$ (the in-plane component and a temperature-dependent bulk material property). A magnetic screening length $\Lambda_S = 2\lambda_{\parallel}^2/d_F$ is defined for a 2D superconductor, where transverse magnetic fields penetrate in the form of quantized fluxoids, or vortices (Pearl, 1964). For a single sheet of a clean superconductor of hypothetical minimum thickness $d_F = d$ (corresponding to the optimal high-$T_C$ compounds) one has $\Lambda_S = m^*c^2/2\pi N_S e^2$, where m* is the effective mass of superconducting carriers and $N_S$ is the sheet (areal) carrier density that is related to volume density $n_S = N_S/d$ (c is speed of light, e is elementary electronic charge); by definition one has $\Lambda_S \gg \lambda_{\parallel}$, where $\Lambda_S$ contains the temperature dependence of $\lambda_{\parallel}^2$.

In principle, fluctuations prevent conventional long range order in two-dimensional systems (Berezinskii, 1971, 1972), including 2D superconductors. However, topological order can exist, which in a superfluid film manifests as bound pairs of thermally excited vortices with counter-flowing circulations that renormalize (or reduce) the superfluid density (Kosterlitz and Thouless, 1973). Such vortex-antivortex pairs dissociate at a theoretical phase transition temperature $T_{KT}$, the Kosterlitz-Thouless transition, where the renormalized superfluid density jumps discontinuously from a universal value to zero (Nelson and Kosterlitz, 1977). Although Kosterlitz and Thouless correctly observe, "This type of phase transition cannot occur in a superconductor …", it was noted by Beasley et al. (1979) that presence of vortex pairs and their dissociation ought to be observable in superconducting thin films within a length scale $r \ll \Lambda_S$, over which magnetic field screening may be neglected. From the perspective of experiment, superconductivity is determined for a physical specimen by either diamagnetic screening or a zero voltage drop in response to an applied magnetic field or electrical current, respectively. In practice the distance $r$ is the lesser between the width of the sample or the least separation of vortex pairs that become dissociated under the force field of a screening or transport current; at finite frequencies the vortex diffusion length also enters (Halperin and Nelson, 1979).

Theory for vortex unbinding was adapted to superconducting films by Halperin and Nelson (1979), where the universal jump at the Kosterlitz-Thouless transition (noting the absence of a true phase transition) was shown to correspond to the relationship $\tilde{\Lambda}_S(T_{KT}) T_{KT} = 2.0$ cm-K, where $\tilde{\Lambda}_S(T_{KT})$ is the renormalized 2D screening length (defined for temperature T approaching $T_{KT}$ from below). The renormalized superfluid density $\tilde{N}_S$, which is proportional to $\tilde{\Lambda}_S^{-1}$, is reduced compared to the microscopic $N_S$ (i.e. at the minimum length scale $r \approx \xi$, where $\xi$ is the superconducting coherence distance), owing to polarization of thermally excited vortex-antivortex pairs, e.g. under the influence of an applied current (the response is analogous to screening of electric fields by dipoles in dielectrics, except that in this case the dielectric constant increases with length scale $r$). Evidence for vortex pair excitation and renormalized superfluid density has been obtained from studies of non-linear current-voltage relationships (Kadin *et al.*, 1983) and dynamical measurements of $\Lambda_S$ (Fiory *et al.*, 1983) in thin (low-$T_C$) superconducting films. The presence of unbound thermally excited vortices at $T > T_{KT}$ yields sheet resistance increasing exponentially with $[-(T - T_{KT})^{-\frac{1}{2}}]$ (Halperin and Nelson, 1979).

The factor by which $\Lambda_S$ is renormalized at $T = T_{KT}$ is the vortex dielectric constant $\varepsilon_v = \tilde{\Lambda}_S(T_{KT}) / \Lambda_S(T_{KT})$, where $\Lambda_S^{-1}(T_{KT}) \propto N_S(T_{KT})$ corresponds to the microscopic superfluid density. Experiments on high-$T_C$ superconductors (where typically $\xi \ll r \ll \Lambda_S$) suggest a value $\varepsilon_v \approx 2$ (Sugimoto *et al.*, 1991; Kim *et al.*, 1991; Matsuda *et al.*, 1992) and we shall employ this value to obtain an estimate of the transition temperature of a thin sheet of a high-$T_C$ superconductor, denoted here as $T_{CS} \approx T_{KT}$. To obtain a model for $T_{CS}$ one also needs the temperature dependence of $\Lambda_S(T)$ or equivalently $N_S(T)$; for this we assume the two-fluid model (Gorter, 1955; Lewis, 1956)

$$N_S(T) = N_{S0} [\, 1 - (T/T_{C0})^4 \,] \,, \qquad (1)$$

which contains the superfluid density at zero temperature $N_{S0}$ and the microscopic transition temperature $T_{C0}$. In Section 3 we apply this model in the form

$$\alpha \equiv T_{CS}/T_{C0} = (2.0 \text{ cm-K}) [\, 1 - (T_{CS}/T_{C0})^4 \,] / \varepsilon_v \Lambda_{S0} T_{C0} \,, \qquad (2)$$

by assuming the bulk-material values for $T_{C0}$ and calculating theoretical estimates of $\Lambda_{S0}$ for thin superconducting layers from the expression

$$\Lambda_{S0} = (m^* c^2 / 2\pi N_{S0} e^2)(1 + \xi/\ell) \,. \qquad (3)$$

Here $\ell$ is the electronic mean free path. The second factor in Equation (3) allows for a possible correction for disorder in thin layer materials (Beasley *et al.*, 1979). Solution of the transcendental Equation (2) determines our estimate of $T_{CS}$ for a single layer of thickness $d$ in terms of the parameter $\alpha$, which is solely a function of $\varepsilon_v \Lambda_{S0} T_{C0}$.

### 3. Unit-cell layers of high-$T_C$ superconductors

The problem that we are considering herein is whether superconductivity can exist, not only within a layer of high-$T_C$ materials of thickness equal to one unit cell, but within the actual-periodicity $d$ as shown in Figure 1 (some of the high-$T_C$ compounds discussed herein, such as $Bi_2Sr_2CaCu_2O_{8+\delta}$ and $La_{2-x}Sr_xCuO_4$, have two formula units per unit cell where the hard-axis lattice parameter of the unit cell is actually $2d$), which contains all of the components (types I and II charge reservoirs) necessary for pairing. In this case, the magnetic screening length at zero temperature $\Lambda_{S0}$ of Equation (3) would be calculated by taking $d$ to be the periodicity as shown in Figure 1, such that $N_{S0}$ is equivalent to $n_{2D}$ from Harshman *et al.* (2011b). Since we will be considering only optimal compounds, we also assume that $\ell \gg \xi$.

Using the measured values of $\lambda_\parallel(0)$, defined as the zero-temperature limit of the bulk magnetic penetration depth (i.e. the basal-plane London penetration depth), from four p-type optimal high-$T_C$ compounds (with $T_{C0}$'s ranging from 10.5 to 93.7 K), and the calculated values of $n_{2D}$, it was found that $m^*/m_0 \approx 1.5$, consistent with the hypothesis of rather small variability in carrier mass (in the superconducting state) and with values close to the bare electron mass $m_0$ (Harshman *et al.*, 2011b). This unique finding was also found to be valid for the one n-type compound tested. By further assuming that $m^* = 1.5\, m_0$ for all high-$T_C$ compounds, it is possible to calculate the corresponding $\Lambda_{S0}$ values for single superconducting sheets of thickness $d$. The results for $N_{S0}$ and $\Lambda_{S0}$ are given in Table 1.

Solutions of the transcendental Equation (2), defining $\alpha \equiv T_{CS}/T_{C0}$, for $\varepsilon_V = 2$ and using the data for $\Lambda_{S0}$ and $T_{C0}$, were obtained for the 32 compounds; results for the ratio $T_{CS}/T_{C0}$, comparing the calculated KT transition temperature $T_{CS}$ to the experimentally determined optimal superconducting transition $T_{C0}$, and $T_{CS}$ are given in Table 1. Frame (a) of Figure 2 shows $\alpha$ (right ordinate axis) and $T_{C0} - T_{CS}$ (left ordinate axis) *vs.* the compound index number from Table 1. The dashed line shows the trend. For the first half of the compounds (i.e., those with the higher $T_{C0}$ values) $\alpha$ is fairly constant, falling off slightly for index numbers above 15; $\alpha$ varies among the compounds from a minimum of 0.51 (No. 24, $La(O_{0.92-y}F_{0.08})FeAs$) to a maximum of 0.95 (No. 11, $HgBa_2CuO_{4.15}$). The plot of $T_{CS} - T_{C0}$ exhibits more scatter and a slightly decreasing trend with decreasing $T_{C0}$ (increasing compound index number).

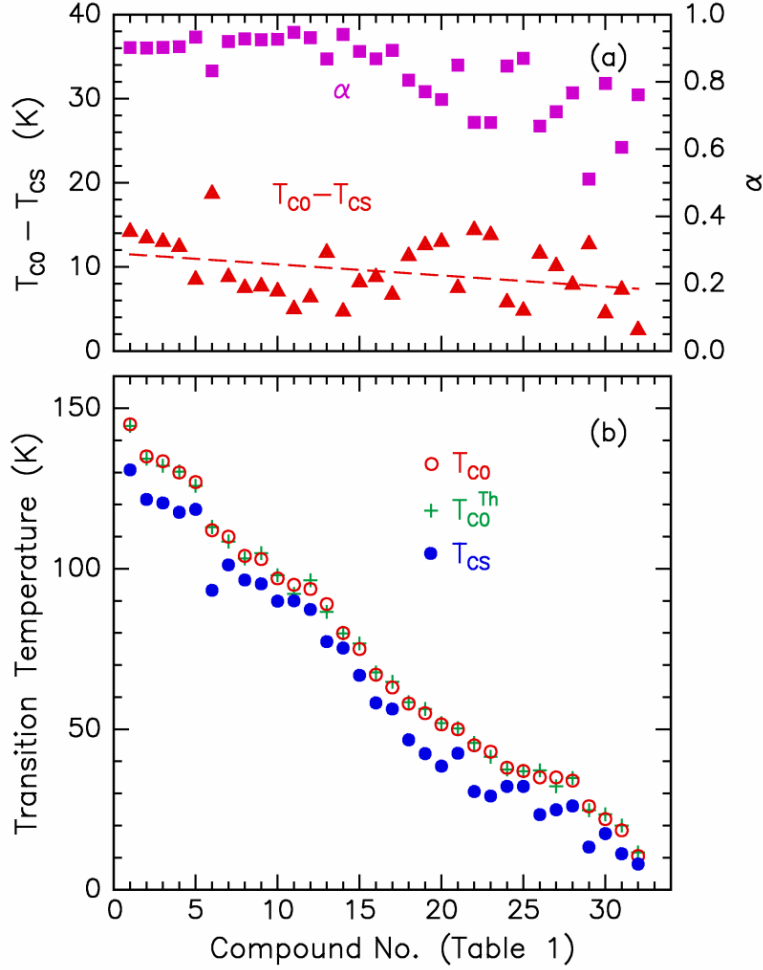

**Figure 2.** Frame (a) presents a graph of $\alpha \equiv T_{CS}/T_{C0}$ (squares, right ordinate) and $T_{C0} - T_{CS}$ (triangles, left ordinate) vs. the compound index number from Table 1. The dashed line denotes the trend. Frame (b) shows transitions temperatures, measured $T_{C0}$, theoretical $T_{C0}^{Th}$ and calculated $T_{CS}$ vs. the compound index number.

The transition temperatures $T_{C0}$, $T_{C0}^{Th}$, and $T_{CS}$ are also compared in Frame (b) where $T_{CS}$ is observed to closely mirror $T_{C0}$ over the full range of compounds.

Figure 3 shows the microscopic $N_S$ (solid curve) and renormalized $\tilde{N}_S(T)$ (dotted curve), assuming $\Lambda_{S0}T_{C0}$ for the "90 K" phase of $YBa_2Cu_3O_{7-\delta}$ from Table 1, both normalized to $N_{S0}$, and assuming two-fluid temperature dependence, normalized to $T_{C0}$, as in Equation (1). The temperature dependence of $\tilde{N}_S$ near $T_{CS}$ reflects the theoretical square-root cusp at the KT transition and the universal jump, which here corresponds to a drop in $\tilde{N}_S/N_S(0)$ from 0.161 to ~0 (Nelson and Kosterlitz, 1977). Data for $N_S(T)/N_S(0)$, shown as open circles, are extracted from $\mu^+$SR (positive-muon spin rotation) measurements of the penetration depth (Pümpin et al., 1990). Note that the $\mu^+$SR data represent bulk microscopic measurements of $N_S(T)$ and therefore closely follow, and indeed validate, the form of the theoretical two-fluid model given in Equation (1) for a constant $m^*/m_0$ (Harshman et al., 2011b). The inset shows the parameter $\alpha$ plotted against $\varepsilon_v\Lambda_{S0}T_{C0}/2$; the curve is the solution of Equation (2) and the symbols correspond to the data in Table 1 (data fall on the curve by definition for $\varepsilon_v = 2$).

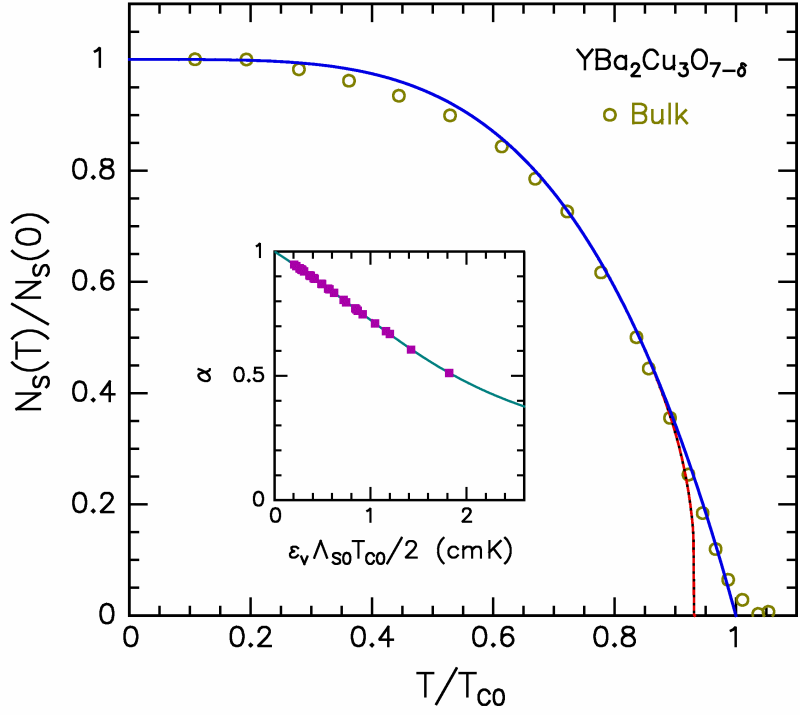

**Figure 3**. $N_S(T)$ normalized to the zero-temperature value $N_S(0)$ as a function of temperature divided by the superconducting transition temperature $T_{C0}$. The solid curve corresponds to the $N_S$ calculated from microscopic theory while the dotted curve is the renormalized $\tilde{N}_S$. The data shown (open circles) are derived from Pümpin et al., (1990). The inset shows the parameter α plotted against $\varepsilon_v \Lambda_{S0} T_{C0}/2$; the curve is the solution of Equation (2) and the symbols correspond to the data in Table 1 (data fall on the curve by definition).

## 4. Experimental tests

In this section we examine several experiments on high-$T_C$ cuprates concerning superconductivity in thin-layer structures of thickness denoted below as $d_F$. Such experiments allow one to compare observed superconducting transitions, denoted as $T_C$, with estimates of transitions from Kosterlitz-Thouless theory, denoted as $T_{CS}$ for such layers. Of particular interest are experiments sensitive to the structure of Figure 1 upon which the microscopic theory for $T_{C0}$ is based. According to this theory an isolated single monolayer is unlikely to be superconducting without an adjacent layer for mediating the superconducting interaction. Specifically, a single type I or type II metallic layer does not support superconductive pairing, owing to the absence of interlayer Coulomb coupling (Harshman et al., 2011b).

There are, of course, notable experimental challenges in preparing ultra-thin high-$T_C$ crystals with layer thickness approaching one unit cell thickness, particularly in reaching the periodicity distance *d* for materials, for example, with two formula units per unit cell. It is an open question (not directly answered by the microscopic theory) as to how one optimally extracts a single-period structure from the bulk, and how to consider the role of the boundary conditions that are introduced (e.g. alteration of charge densities in outer layers). From a practical perspective, the simplest structures are thin films deposited on a substrate where the main considerations are forming an epitaxial and stoichiometric phase of the desired

high-$T_C$ compound (i.e. with the crystal *c*-axis oriented perpendicular to the plane of the substrate). In the following we consider data from published experiments on ultra-thin layers of the cuprate superconductors, $Bi_2Sr_2CaCu_2O_{8+\delta}$, $La_{1.56}Sr_{0.44}CuO_4$, and $YBa_2Cu_3O_{7-\delta}$, which were prepared by three different film deposition techniques. For each compound the value $m^* = 1.5\ m_0$ employed in the calculations for $\Lambda_{S0} = m^*c^2/2\pi N_{S0}e^2$ in Table 1 has been validated by noting that the relation $\Lambda_{S0} = 2\lambda_{\parallel}^2(0)/d$ is obeyed, corresponding to 1% reproducibility in $m^*$ among the compounds and 10% experimental uncertainty in $\lambda_{\parallel}(0)$ (Harshman *et al.*, 2011b).

**4.1 $Bi_2Sr_2CaCu_2O_{8+\delta}$**

In the experiments of Sugimoto *et al.* (1991) and those of Saito and Kaise (1998) ultra-thin films of $Bi_2Sr_2CaCu_2O_{8+\delta}$ (Bi2212) of various thickness $d_F$ were grown on MgO substrates. The thinnest films prepared by both groups contained the targeted Bi2212 phase. X-ray diffraction measurements found that films thicker than about 4 nm contained various admixtures of the $Bi_2Sr_2Ca_2Cu_3O_{10+\delta}$ phase (Bi2223). For the Bi2212 compound one has $d = 1.5445$ nm (Table 1) and the unit cell corresponds to a film 3.089-nm thick. Transition temperatures $T_C$ were estimated from the data presented in these works by extrapolating the steepest part of the resistivity-*vs.*-temperature curves (noting that KT theory predicts a divergent slope at $T_{KT}$, this method helps to compensate for inhomogeneous transitions). Data for $T_C$ obtained from these two studies are plotted as functions of $d_F/d$ in Figure 4, where "A" denotes films from the work of Saito and Kaise (1998) and "B" from Sugimoto *et al.* (1991). The plot excludes the thickest film from each work that failed to grow as single-phase Bi2212. Films in the work of Sugimoto *et al.* for $d_F \geq 3.5$ nm show well-defined superconducting transitions, whereas films for $d_F \leq 2.2$ nm are non-metallic and non-superconducting; a film for $d_F = 2.7$ nm shows evidence for superconducting transitions at two temperatures (two open-circle points plotted for "B" in Figure 4). Films for $d_F \geq 2.7$ nm fall into the classification of good metals (T > 15K), since their temperature coefficients of resistivity are positive. In the later work of Saito and Kaise (1998) metallic and superconducting films were obtained for a smaller minimum thickness of $d_F = 2.0$ nm (a 1.0-nm film was found to be almost insulating, $T_C \sim 0$) and the values for $T_C$ are higher than found in the earlier work. Both works show a trend of $T_C$ increasing with film thickness, part of which may be attributed to the higher microscopic $T_C$ of the Bi2223 phase.

Normal state resistivities $\rho_F$ obtained by extrapolation of the temperature dependence to $T_C$ (or at the lowest measured temperature for the non-superconducting films) were used to calculate the quantity $k_F\ell$ (product of the in-plane Fermi wave vector and mean free path) by modeling the film as a stack of two-dimensional sheet resistances and taking the effective number of conducting sheets to be given by $d_F/d$. From theory of metallic transport in 2D, one has,

$$k_F\ell = (h/e^2)\ d\ \rho_F^{-1}\ , \qquad (4)$$

where h is Planck's constant. The results for $k_F\ell$ are shown in the inset in Figure 4. The data show that the onset of superconductivity (and metallic behaviour in the normal state) coincides with a jump in $k_F\ell$ from <<1 for the marginally superconducting films to $\geq 8$ for the superconducting films. This is similar to the transition from an insulator (in the limit of zero temperature) to a metallic superconductor that has been observed in very thin films of conventional superconducting metals and is associated with carrier localization and the pair-breaking effects of disorder (Hebard and Paalanen, 1984; Graybeal and Beasley, 1984; Haviland *et al.*, 1989).

The solid curve in Figure 4 denoted $T_{CS}$ shows the KT transition temperature calculated from the parameters in Table 1 ($\Lambda_{S0}$ = 55.6 μm, $T_{C0}$ = 89 K) scaled to a layer of thickness $d_F$, assuming that the entire entity of the films in each work acts as a single 2D superconductor. This implicitly assumes that the microscopic $T_{C0}$ in the film is unchanged from the bulk (one notes that $T_{C0}$ varies with oxygen doping δ and annealing, and may vary among these works). If one assumes instead that the film comprises multiple layers of 2D superconductors spaced a distance $d$ apart, as has been indicated from KT transitions in bulk crystals of $Bi_2Sr_2CaCu_2O_{8+δ}$ (e.g. Martin *et al.*, 1989) then the applicable transition is $T_{CS}$ = 77.3 K, as given in Table 1 and denoted by the star symbol on the curve. This value is bracketed by the $T_C$ results for the thinnest superconducting films as obtained in the two works: 67 K and 86 K for the two 2.0-nm films of Saito and Kaise (1998) and $T_C$ ~ 69 K for the 2.7-nm film in the work of Sugimoto *et al.* (1991). Film quality evidently improves with thickness as indicated by the trend for $k_F\ell$ to increase with $d_F/d$. A 7.0-nm film studied by Saito and Kaise (1998), which is mostly the Bi2223 phase (essentially the same as compound No. 6 in Table 1 with $T_{C0}$ = 110 K and $d$ = 1.8541 nm), exhibits $T_C$ = 107 K (by extrapolation). We obtain the estimate $T_{CS}$ = 107.3 K by modeling the 7.0-nm film as comprising pure Bi2223 phase with $T_{C0}$ =110 K.

The fact that superconductivity is abruptly suppressed below about 2.0 nm or ~1.3$d$, is consistent with expectations of our Coulombic mediation pairing model, which requires crystals of minimum thickness in the range $d \leq d_F \leq 2d$ at the least.

In their study of $Bi_2Sr_2(Ca_xY_{1-x})Cu_2O_8$ superlattice films (~60 nm total thickness) Kanai *et al.* (1990) compared various superlattices comprising multi-layers of superconducting $Bi_2Sr_2(Ca_{0.15}Y_{0.85})Cu_2O_8$ interleaved with multi-layers of semiconducting $Bi_2Sr_2(Ca_{0.5}Y_{0.5})Cu_2O_8$ spacer layers. Transition temperatures (from midpoints in resistance-*vs.*-temperature data) were ~ 65 K for films containing $Bi_2Sr_2(Ca_{0.15}Y_{0.85})Cu_2O_8$ layers of thickness 6.0, 12.0, and 60 nm; the thickness independence is consistent with bulk-like behavior. A slightly depressed $T_C$ ~ 63 K was found for superconducting layers of 3.0-nm ($\approx 2d$) thickness. Although the Y-doped superconductor under study is not the optimum $Bi_2Sr_2CaCu_2O_8$ of Table 1, the onsets of the superconducting transitions at ~ 90 K are comparable to $T_{C0}$ = 89 K. This experiment thus also indicates that nearly bulk-like superconductivity is contained in a superconducting layer of one-unit cell thickness.

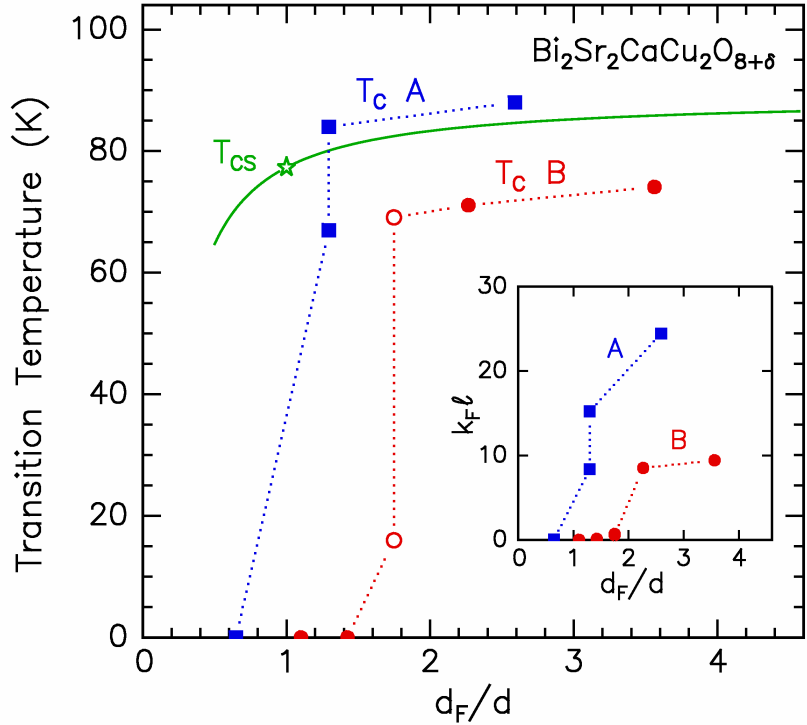

**Figure 4**. The transition temperature $T_C$ and $k_F\ell$ (inset), both plotted as a function of $d_F/d$ for thin-film $Bi_2Sr_2CaCu_2O_{8+δ}$ samples; The data denoted as "A" (blue squares) correspond to Saito and Kaise (1998) and "B" (red circles) are from Sugimoto *et al.* (1991). The solid curve is the theoretical calculation of $T_{CS}$ *vs.* $d_F/d$ with $T_{C0}$ = 89 K; the green star corresponds to $d_F/d$ = 1.

## 4.2 $La_{2-x}Sr_xCuO_4$

In a recent experiment conducted by Logvenov *et al.* (2009) bilayer $La_{1.56}Sr_{0.44}CuO_4/La_2CuO_4$ films were grown by molecular beam epitaxy on a $LaSrAlO_4$ substrate, with the thickness of each half of the bilayer set at three unit cells (six formula units) and with La-Sr-O and La-O monolayers straddling the defined interface. The Sr-doped side is over-doped, metallic and non-superconducting; the undoped side is nominally non-metallic and non-superconducting. From considerations of diffusion of Sr across the interface, Logvenov *et al.* (2009) determine that a superconducting region of unit-cell thickness with varying Sr doping levels is produced on the $La_2CuO_4$ side (corresponding to $d_F \approx 2d$ in our notation). The authors prepared a series of samples with 3% Zn substituting for Cu in a single monolayer designated by an index number n (n = –6, …, –1 in the side grown with Sr doping and n = 1, …, 6 in the side nominally without Sr). Samples containing Zn-substituted $CuO_2$ monolayers far from the interface (|n| = 5 … 6) show $T_C \approx 33 - 35$ K and are similar to samples without Zn ($T_C$ = 34 K). Samples with Zn in layer n=2 (second $CuO_2$ monolayer from the interface) yield lowest $T_C$ ~ 17 – 23 K, similar to depressed $T_C$ ~ 18 K of bulk crystals containing 3% Zn. However, Zn substitution in the two layers adjacent to n=2 causes partially depressed transitions: $T_C$ ~ 29 – 35 K for n=1 and $T_C$ ~ 29 – 30 K for n=3. This pattern of variation of $T_C$ with the Zn-substituted layer is therefore consistent with a superconducting region of unit-cell thickness, or $d_F \approx 2d$.

While the experimental reproducibility is somewhat limited, we have analyzed data for $\Lambda_S(T)$ in Logvenov *et al.* (2009) to obtain experimental values for $\Lambda_S(0)$ (compared below to the theoretical $\Lambda_{S0}$ of Table 1) from the lowest temperature measured (5 K) and $T_C$ from the largest $\Lambda_S$ measured (1.25 cm); the results are reproduced in Figure 5 as filled blue circles for the n ≠ 2 samples, denoted MBE-Zn(n), and as filled blue triangles for three n = 2 samples, denoted MBE-Zn(2). For comparison, data for bulk $T_C$ *vs.* $\Lambda_S(0) = 2\lambda_\parallel^2(0)/d$, as determined from measurements of $\lambda_\parallel(0)$ in crystalline $La_{2-x}Sr_xCuO_4$ (Tallon *et al.*, 2003) and $d$ = 0.66029 nm from Table 1, are shown as red squares, where x increases from right to left; points to the right of maximum $T_C$ are underdoped (x < 0.163) and points to the left are overdoped (x > 0.163); the dotted curve follows the smooth trend. Note that the overdoped bulk samples indicate a lower limit on $\Lambda_S(0)$ of 119 μm. Also shown is the theoretical $T_{CS}$ *vs.* $\Lambda_S(0)$ from KT theory (with $\varepsilon_v$ = 2), plotted for $d_F/d$ = 1 (solid curve) and $d_F/d$ = 2 (dashed curve). The star symbol marks the theoretical ($\Lambda_{S0}$, $T_{CS}$) datum point for the optimum compound $La_{1.837}Sr_{0.163}CuO_4$ in Table 1; the theoretical value $\Lambda_{S0}$ = 150.8 μm is in excellent agreement with experimental value $\Lambda_S(0)$ = 156 μm determined from $\lambda_\parallel(0)$ = 0.227 μm and the measured $d$ for this compound.

The thickness $d_F$ may be determined from examination of the MBE film data in Figure 5. Notice that all of the MBE-Zn(n) data (i.e. Zn substituted $La_{2-x}Sr_xCuO_4$, n ≠ 2), represented by the circles, fall to the left of the theoretical value marked by the star symbol on theoretical curve. Further, all of the MBE-Zn(n) data also fall to the left of the bulk overdoped $La_{2-x}Sr_xCuO_4$ data (red squares). Both of these trends imply $d_F > d$. All but one (a circle) of the MBE data points lie below the theory curve for $d_F = d$; most of the MBE data are obviously rather close to the theory curve for $d_F = 2d$.

Since Zn substitution and non-optimal Sr doping both depress $T_C$ by essentially equivalent mechanisms of disorder, one may compare the bulk and MBE film data to ascertain estimates of $d_F$ purely from experimental data. This is accomplished by applying the expression $d_F = 2\lambda_\parallel^2(0)/\Lambda_S(0)$, using the MBE-sample data for $\Lambda_S(0)$ and the bulk-crystal data for $\lambda_\parallel(0)$ that correspond to the same $T_C$ values (interpolating bulk-crystal data as needed). The only remaining unknown is the Sr doping of the superconducting layers in the MBE films, which introduces an uncertainty as to which branch of the bulk-crystal data to use, overdoped or underdoped (although the trend of the MBE data appears to closely mimic the underdoped branch of the bulk-crystal data). We show the results for these estimates of $d_F$ in the inset to Figure 5, noting that the corresponding MBE data in the main figure tend to cluster in three groups (seven circles, three circles, and three triangles). Each data group is represented in the inset by a datum symbol, where horizontal bars denote the range of transition temperatures within the group; the vertical bars show the standard deviation, which includes the scatter in the data and uncertainties in doping status (allowing for either overdoped, underdoped or a mixture of the two) and the expectation that $d_F \geq d$. Averaging these results one obtains $d_F/d = 2.5 \pm 1.4$; it is evident, however, that $d_F/d$ trends to 2 for the MBE-Zn(2) samples and to greater than 2 for the MBE-Zn(n) samples that are closest to being optimum.

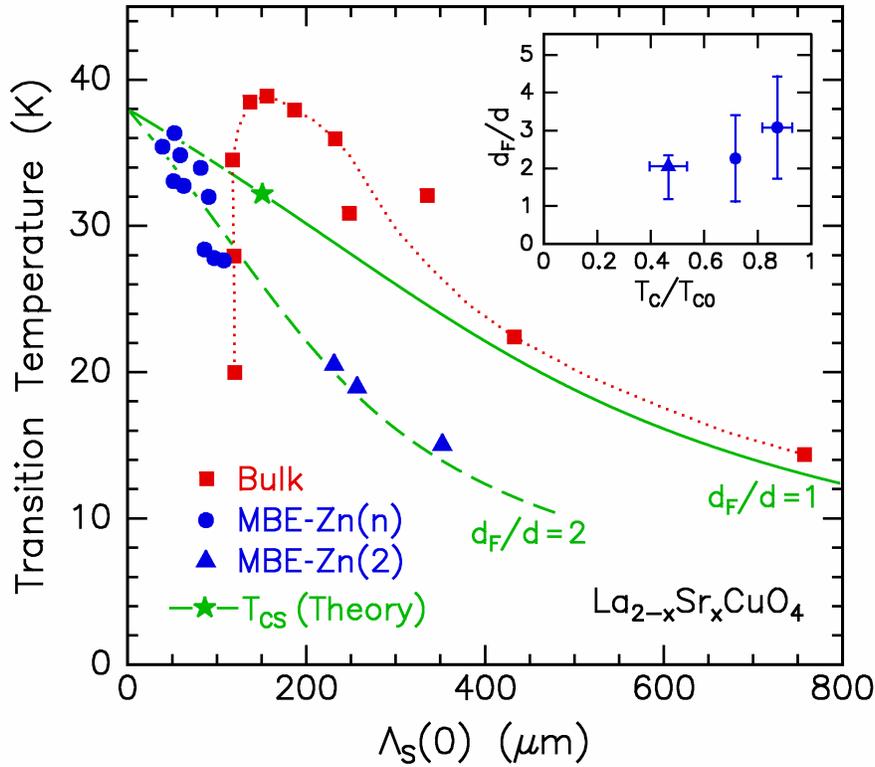

**Figure 5**. The transition temperature ($T_C$ for data, $T_{CS}$ for theory) plotted against the zero-temperature screening length $\Lambda_S(0)$ for bulk crystals (red squares, Tallon *et al.*, 2003) and MBE-grown films (blue circles and triangles, Logvenov *et al.*, 2009). The dotted line is a guide to the eye and the solid and dashed green curves are theoretical functions of $T_{CS}$ *vs.* $\Lambda_S(0)$ corresponding to $d_F/d = 1$ and $d_F/d = 2$, respectively; the green star marks the theoretical $T_{CS}$ at $\Lambda_S(0) = \Lambda_{S0}$ for $La_{1.837}Sr_{0.163}CuO_4$. Inset: calculated $d_F/d$ *vs.* reduced $T_C/T_{C0}$ for the MBE-grown data.

This experiment, therefore, demonstrates that Zn substitution at n=2 clearly affects the superconductivity over at least a full unit cell (containing two $CuO_2$ monolayers), and not at a single $CuO_2$ monolayer as presumed by Logvenov et al. (2009). It is important to understand the pronounced depression in $T_C$ for Zn substitution in the n=2 $CuO_2$ monolayer, while taking into consideration the partial and non-negligible effects on $T_C$ for Zn in either of the adjacent n=1 and n=3 monolayers. Within the periodicity distance $d$ for $La_{2-x}Sr_xCuO_4$ there are two La/SrO-La/SrO type I ionic bilayers separated by and straddling a $CuO_2$ type II ionic monolayer. Our model of high-$T_C$ superconductivity is based on this structure (generalized in Figure 1, lower panel), and is entirely consistent with and validated by the results of this experiment. One readily recognizes that disordering the type II layer by Zn substitution affects a region of thickness $2d$, not just simply $d$, since this modifies the Coulomb interactions between the $CuO_2$ monolayer and the two La/SrO-La/SrO bilayers on either side. In our model, superconductivity is degraded by disorder within any part of the complete type I / type II structure. This interpretation also recognizes that a La/SrO-La/SrO bilayer in $La_{2-x}Sr_xCuO_4$ is essentially one unit, since the constituent La/SrO monolayers are intimately connected. The significant, albeit smaller, depressing effect on $T_C$ for Zn substitution in the two $CuO_2$ layers at the boundaries of the $d_F = 2d$ superconducting structure shows that maintaining bulk boundary conditions is also important for the superconductivity in thin layers.

The disorder behavior observed in the Zn-doped bilayer $La_{1.56}Sr_{0.44}CuO_4/La_2CuO_4$ films is similar to what is observed for $Bi_2Sr_2CaCu_2O_{8+\delta}$ films discussed in section 4.1 above. Mean-free-path information for the MBE samples is unavailable because the corresponding resistance data were presented in normalized form, so a similar analysis cannot be applied. However, as a consistency check in this case, one may observe from Figure 5 that the underdoped bulk data, as well as the MBE data with the lowest transition temperatures produced by the Zn doping, both tend toward the KT theory at large $\Lambda_S(0)$. This is an expected result, owing in part to the concomitantly increasing effective mass anisotropy with decreasing Sr doping x (Shibauchi et al., 2004).

### 4.3 $YBa_2Cu_3O_{7-\delta}$

When grown directly on substrates (e.g. $SrTiO_3$ or MgO) ultra-thin layers of $YBa_2Cu_3O_{7-\delta}$ tend to nucleate in the form of two-dimensional islands and grow in unit-cell blocks, which is associated with the thermodynamics of anisotropic surface energy and kinetics (Pennycook et al., 1992). Largely for this reason, studies of superconductivity in ultra-thin $YBa_2Cu_3O_{7-\delta}$ have focused on multiple-layer structures in which superconducting $YBa_2Cu_3O_{7-\delta}$ layers of various thicknesses are interleaved with or clad by lattice-matching non-superconducting layers (Matsuda et al, 1992; Cieplak et al., 1994; Repaci et al., 1996). Resistive transitions indicate a minimum thickness for the onset of superconductivity that corresponds very closely to one unit cell, $d_F \approx d = 1.16802$ (Cieplak et al., 1994).

As discussed further in Section 5, the non-superconducting (semiconducting) buffer layers chosen for these studies have been $PrBa_2Cu_3O_{7-\delta}$ or $(Pr,Y)Ba_2Cu_3O_{7-\delta}$, both of which can actually be superconductors with bulk $T_{C0}$ nearly the same as for $YBa_2Cu_3O_{7-\delta}$. It is also known that Pr readily diffuses in $YBa_2Cu_3O_{7-\delta}$ and produces Pr-on-Ba-site defects that induce disorder sufficient to depress or destroy the superconductivity (Harshman et al., 2008). This explains why the Pr-doped cladding layers are non-superconducting. The nominally pure $YBa_2Cu_3O_{7-\delta}$ layers are therefore likely to be diffusion-doped with Pr, as indicated by the depressed transition temperatures and depressed conductance at the superconducting/semiconducting interfaces (Cieplak et al., 1994).

Applying the same extrapolation analysis of resistive transitions used in Section 4.1 to data obtained for nominally one-unit-cell $YBa_2Cu_3O_{7-\delta}$ structures reported in Repaci et al. (1996), Matsuda et al. (1992), and Cieplak et al. (1994), one obtains $k_F\ell \approx 2.7$, $12 - 22$, and 20, corresponding to $T_C$ of 37, $60 - 66$, and 20 K, for the three studies, respectively. This assumes two-dimensional conductivity in a layer of thickness $d_F = d$ (in some models the conducting sheets are presumed to be individual $CuO_2$ or BaO layers, for which the calculated $k_F\ell$ would be one-half as large). The absence of systematic correlation between $T_C$ and $k_F\ell$ (or sheet resistance) among the three experiments appears to be symptomatic of irreproducibility in the Pr auto-doping effects that degrade both measures of sample quality. Based on observed onsets of resistive transitions, Matsuda *et al.* (1992) suggest a microscopic $T_C$ of 78 – 82 K, which is below the bulk $T_{C0}$ (93.7 K) and the estimated $T_{CS}$ (87.3 K) from Table 1.

Because of uncertainties introduced by the Pr-doping techniques, data for transition temperatures and magnetic screening lengths from these experiments are insufficient for making quantitative tests of theory for either $T_{C0}$ or $T_{CS}$. However, these experiments do indicate that a minimum thickness of one unit cell is required for superconductivity ($d_F \approx d$). This result is consistent with our model structure (in this case exactly like Figure 1, upper panel), where the type I structure is BaO-CuO-BaO and the type II structure is $CuO_2$-Y-$CuO_2$, both comprising three monolayers. A unit cell is therefore the minimum thickness that contains all the required components to support superconductivity in a stand-alone structure. The logical interpretation is that both sets of interacting layers, $CuO_2$/BaO, lie within the structure and either Y or CuO spacer/doping monolayers are at the boundaries. Inability to satisfying the boundary conditions provides the natural explanation for the experimental observation that superconductivity is absent in films for $d_F < d$ (Cieplak *et al.*, 1994; Rüfenacht *et al.*, 2003). Further, the step-wise behavior of $T_C$ and sheet conductivity corresponding to integer ratios of $d_F/d$ (Cieplak *et al.*, 1994) suggests that a fractional unit cell of thickness less than $d$ is non-superconducting and non-metallic.

## 5. Discussion

Moving closer to a better understanding of the detailed physical nature of the high-$T_C$ pairing mechanism is the ultimate goal of this and other work. Recently, it was shown unequivocally that the superconductivity in high-$T_C$ compounds is governed by Coulomb interactions (Harshman *et al.*, 2011b), with the model high-$T_C$ structure given in Figure 1; initial validation of the pairing model developed therein was obtained by predicting the transition temperature values $T_{C0}$ from 31 different optimal compounds from 5 different high-$T_C$ sub-families. In order to take a closer look at the microscopic mechanism driving the superconductivity in such diverse 2D compounds, we consider the problem in thin films in the limit of KT theory.

In our application of the KT theory to layers of thickness equal to the crystalline periodicity $d$, and utilizing the $N_{S0}$ values and other parameters from Harshman *et al.* (2011b), we calculate the superconducting sheet transition temperature $T_{CS} = \alpha T_{C0}$ ($\alpha < 1$) for 32 high-$T_C$ compounds, including the transuranic $PuCoGa_5$. By adopting a two-fluid form for $\lambda_\parallel(T)$, $\alpha$ was determined to be on average $0.83 \pm 0.11$ (see Figures 2 and 3).

Most importantly, experimental studies of thin film $Bi_2Sr_2CaCu_2O_8$ (Sugimoto *et al.*, 1991; Saito and Kaise, 1998), $La_{2-x}Sr_xCuO_4$ (Logvenov *et al.*, 2009) and $YBa_2Cu_3O_{7-\delta}$ (Cieplak *et al.*, 1994), where $d_F$ approaches $d$, all place an important restriction or boundary condition, $d_F \geq d$, for these samples to be superconducting. This boundary condition is also implied since the end growth surfaces of $Bi_2Sr_2CaCu_2O_{8+\delta}$ and $YBa_2Cu_3O_{7-\delta}$ films are known to terminate on a (single) BiO and CuO layer, respectively (Wen *et al.*, 1995). Thus, it is clear that for superconductivity to be induced, one requires (at least) a complete set of type I / type II charge reservoir structures as depicted in Figure 1, consistent with our pairing theory. Unfortunately, the experiment of Logvenov *et al.* (2009) with $d_F \sim 2d$ appears to have insufficient resolution for determining whether half-cell $d_F = d$ or single-$CuO_2$ structures of $La_{2-x}Sr_xCuO_4$ are superconducting.

The conceptual designs of the unit-cell $YBa_2Cu_3O_{7-\delta}$ experiments (Terishima *et al.*, 1991; Matsuda *et al*, 1992; Cieplak *et al*., 1994; Repaci *et al.*, 1996), while elegant, partially failed in practice owing to Pr diffusion from the semiconducting Pr-doped $YBa_2Cu_3O_{7-\delta}$ cladding. These experiments were nevertheless able to indicate a minimum thickness for the onset of superconductivity at $d_F \approx d$. A very large number of experiments have been carried out under the misconception that Pr only doped the Y site in $YBa_2Cu_3O_{7-\delta}$, which is patently untrue; Pr doping of this compound actually induces $Pr^{+3}$-on-$Ba^{+2}$-site defects, which are solely responsible for the reduction and eventual complete suppression of superconductivity. Properly grown $PrBa_2Cu_3O_{7-\delta}$ is superconducting with $T_{C0} \sim 90$ K (Zou et al., 1997; Zou et al., 1998; Shukla et al., 1999). Like the isotope effect behavior which was dealt with earlier by Harshman et al. (2008), many experiments conducted on Pr-doped $YBa_2Cu_3O_{7-\delta}$, may have to be reinterpreted with Pr diffusion properly considered before meaningful information can be extracted. For example, Repaci et al. (1996) had utilized a sample with the lowest $k_F\ell$, indicating the most disordered sample had been studied in that work; their observation of residual low-temperature resistance was interpreted as the absence of a KT transition. This experiment has stimulated an alternative dynamical scaling analysis of these and similar data by Pierson et al. (1999), who obtain a fitted KT temperature of 17.6 K; however, further analysis by Strachan et al. (2003) of data from Repaci et al. (1996) and Matsuda et al. (1992) found that such scaling theories (e.g. applied to non-linear current-voltage curves) have intrinsic flexibility and arbitrariness and therefore have not been put to a critical test.

Much research in the high-$T_C$ field is conducted on non-optimal compounds, which are usually characterized by reduced Meissner fractions and broadened transition widths; phase separation eventually occurs in many (if not most) of these compounds if the stoichiometry strays too far from optimum. As an example one may consider the kinetic inductance measurements of $\Lambda_S(T)$ for ultra-thin films of $La_{2-x}Sr_xCuO_4$, where samples with $T_C \sim 8$ K are found to yield microscopic penetration depths scattered over a wide range (e.g. for three films, $\lambda_\parallel(0) = 0.535$, 0.760, and 2.3 μm), apparently varying with MBE structure (Rüfenacht et al., 2003). As an example of reduced Meissner effect, another study of an ultrathin film with $T_C \sim 8$ K was observed to have $\Lambda_S(0) \sim 1$ cm (Rüfenacht et al., 2006); this implies a datum ~12 times off scale to the right in Figure 5 and disagrees with KT theory, i.e. $T_{CS} \sim 1$ K. While microscopic interpretations for these anomalies may be of some interest, the more likely scenario is that modeling such films as a uniform two-dimensional sheet is inadequate.

As one moves away from optimum, electron-phonon scattering is also observed to increase rapidly, as evidenced by the increasing mass-exponent of the oxygen isotope effect (Harshman et al., 2008) with non-optimization (in optimal compounds this exponent is typically vanishingly small). Since the intrinsic superconducting parameters such as $\lambda_\parallel(0)$ and $\xi_0$ (Pippard coherence distance) are only defined in the optimal case, it is unclear what measurements on non-optimal compounds tell us about the superconducting state except that it can be destroyed by introducing enough disorder. As such effects are difficult to quantify, the microscopic theory tested herein is concerned only with optimal, or near-optimal high-$T_C$ compounds (e.g. it cannot yet be used to predict the reduced $T_C$ values for off-stoichiometric materials).

## 6. Conclusion

The discovery and validation of the Coulombic nature of the high-$T_C$ pairing mechanism, with the optimal transition temperature given by $k_B T_{C0} = \beta (N_{int})^{1/2} \zeta^{-1}$ (Harshman et al., 2011b), has prompted further investigations into the precise nature of the pairing mechanism; specifically it entails seeking further evidence of the essential roles played by the type I and type II charge reservoir components (all contained within periodicity $d$) in forming the superconducting condensate and mediating pairing through interlayer Coulombic interactions. To this end, we have considered the high-$T_C$ problem in the case of ultra-thin crystals having thicknesses $\sim d$. Sheet transition temperatures $T_{CS} = \alpha T_{C0}$ ($\alpha < 1$) are determined from Kosterlitz-Thouless theory for thirty-two cuprate, ruthenate, rutheno-cuprate, iron pnictide, organic, and transuranic compounds. Assuming parameters derived elsewhere (Harshman et al., 2011b) and adopting a two-fluid form for $\lambda_\parallel(T)$, $\alpha$ was found to be on average $0.83 \pm 0.11$. The results are summarized in Table 1, and Figures 2 and 3, and provide a tangible connection between microscopic theory (Harshman et al., 2011b) and the behaviour of superconducting thin films in the limit $d_F = d$.

Calculations of $T_{C0}$ from the microscopic theory of Harshman *et al.* (2011b) and $T_{CS}$ from KT theory were also shown to be consistent with several experiments on thin crystal structures (thin films) of thickness $d_F$ approaching $d$. Of particular interest is the research reported on $Bi_2Sr_2CaCu_2O_{8+\delta}$ thin films (Sugimoto *et al.*, 1991; Saito *et al.*, 1998), see Figure 4, which shows that the superconductivity is suppressed for layer thickness $d \leq d_F \leq 2d$. Measurements of thin film structures of $YBa_2Cu_3O_{7-\delta}$ also indicate that a minimum thickness of one unit cell (i.e., $d_F \approx d$) is required to induce superconductivity (Cieplak *et al.*, 1994); this minimum requirement is also consistent with observation that the end surfaces of $Bi_2Sr_2CaCu_2O_{8+\delta}$ and $YBa_2Cu_3O_{7-\delta}$ films terminate on a (single) BiO and CuO layer, respectively (Wen *et al.*, 1995). Clearly the periodicity $d$ contains all the components (complete type I and type II reservoirs) necessary to support superconductivity. Experiments with selective monolayer Zn-substitution in $La_{2-x}Sr_xCuO_4$ multilayer films (Logvenov *et al.*, 2009) have shown that a bulk-like superconducting condensate can exist in a layer $d_F \sim 2d$, containing approximately two formula units. Whether high-$T_C$ superconductivity may exist in single-formula-unit structures, e.g. $d_F = d$, or even down to a single $CuO_2$ monolayer as has been claimed, remains unresolved by this method for this compound. In our picture, disorder introduced to any component of the complete type I / type II structure degrades the superconductivity. Assigning the superconducting condensate in the (type I) La/SrO layers and the mediating carriers to the (type II) $CuO_2$ planes presents a scenario that is consistent with of all of the experimental evidence discussed.

It is interesting that in the experiments considered for this study and others, the practicality of a minimum crystal thickness appears to correspond to no less than one formula unit. The evident implications are that a complete formula unit is the minimum amount of material required for replicating the bulk-like structure in an ultra-thin crystal as well as for providing the two interacting charge reservoirs of high-$T_C$ superconductors. According to our microscopic theory, an isolated conducting sheet without an adjacent charge layer to mediate the pairing interaction would not be superconducting; this inference from theory is thus far borne out by experiments on ultra-thin crystals.

### Acknowledgements

The authors wish to thank Prof. Ravindra for calling to our attention this opportunity to participate in the inaugural issue of Emerging Materials Research. The year of this writing marks the 100[th] anniversary of the discovery of superconductivity by Heike Kamerlingh Onnes (Onnes, 1911; van Delft and Kes, 2010). We are also grateful for the support of Physikon Research Corporation (Project No. PL-206) and the New Jersey Institute of Technology. Publication of this work has appeared (Harshman2012).

Cieplak M Z, Guha S, Vadlamannati S, Giebultowicz T and Lindenfeld P (1994) Origin of the $T_C$ depression and the role of charge transfer and dimensionality in ultrathin $YBa_2Cu_3O_{7-\delta}$ films. *Phys. Rev. B* **50(17)**: 12876-12886.

Clem J R (1991) Two-dimensional vortices in a stack of thin superconducting films: A model for high-temperature superconducting multilayers. *Phys. Rev. B* **43(10)**: 7837-7846.

de la Cruz C, Huang Q, Lynn J W, Li J, Ratcliff W II, Zarestky J L, Mook H A, Chen G F, Luo J L, Wang N L and Dai P (2008) Magnetic order close to superconductivity in the iron-based layered $LaO_{1-x}F_xFeAs$ systems. *Nature* **453(7197)**: 899-902.

Fiory AT, Hebard A F and Glaberson W I (1983) Superconducting phase transitions in indium/indium-oxide thin-film composites. *Phys. Rev. B* **28(9)**: 5075-5087.

Fiory A T, Hebard A F, Mankiewich and Howard R E (1988) Renormalization of the mean-field superconducting penetration depth in epitaxial $YBa_2Cu_3O_7$ films. *Phys. Rev. Lett.* **61(12)**: 1419-1422.

Gorter C J (1955) Chapter I The Two Fluid Model for Superconductors and Helium II. *Progress in Low Temperature Physics* (Elsevier) **1**:1-16

Graybeal J M and Beasley M R (1984) Localization and interaction effects in ultrathin amorphous superconducting films. *Phys. Rev. B* **29(7)**: 4167-4169.

Gurvitch M and Fiory A T (1987) Resistivity of $La_{1.825}Sr_{0.175}CuO_4$ and $YBa_2Cu_3O_7$ to 1100 K: Absence of saturation and its implications. *Phys. Rev. Lett.* **59(12)**: 1337-1340.

Gurvitch M, Fiory A T, Schneemeyer L F, Cava R J, Espinosa G P and Waszczak J V (1988) Resistivities of ceramic and single-crystalline superconducting oxides to 1100 K: What do they tell us? Physica C **153-155(3)**: 1369-1370.

Halperin B I and Nelson D R (1979) Resistive transition in superconducting films. *J. Low Temperature Phys.* **36(5/6)**: 599-616.

Harshman D R and Mills A P Jr. (1992) Concerning the nature of high-$T_C$ superconductivity: Survey of experimental properties and implications for interlayer coupling. *Phys. Rev. B* **45(18)** 10684-10712.

Harshman D R, Dow J D and Fiory A T (2008) Isotope effect in high-$T_C$ superconductors. *Phys. Rev. B* **77(2):** 024523(1-9).

Harshman D R, Dow J D and Fiory A T (2009) Reply to "Comment on 'Isotope effect in high-$T_C$ superconductors' ". *Phys. Rev. B* **80(13):** 136502(1-6).

Harshman D R, Dow J D and Fiory A T (2011a) Coexisting holes and electrons in high-$T_C$ materials: Implications from normal state transport. *Philos. Mag.* **91(5)**: 818-840.

Harshman D R, Fiory A T and Dow J D (2011b) Theory of high-$T_C$ superconductivity: Transition temperature. *J. Phys.: Condens. Matter* (to be published).

Harshman DR and Fiory AT (2012) High-$T_C$ superconductivity in ultra-thin crystals: Implications for microscopic theory *Emerging Materials Research* (DOI: 10.1680/emr.11.00001, in press).

Haviland D B, Liu Y and Goldman A M (1989) Onset of superconductivity in the two-dimensional limit. *Phys. Rev. Lett.* **62(18)**: 2180-2183.

Hebard A F and Fiory A T (1980) Evidence for the Kosterlitz-Thouless transition in thin superconducting aluminum films. *Phys. Rev. Lett.* **44(4)**: 291-294.

Hebard A F and Paalanen M A (1984) Pair-breaking model for disorder in two-dimensional superconductors. *Phys. Rev. B* **30(7)**: 4063-4066.

Hoen S, Creager W N, Bourne L C, Crommie M F, T.W. Barbee T W III, Cohen M L, Zettl A, Bernardez L and Kinney J (1989) Oxygen isotope study of $YBa_2Cu_3O_7$. *Phys. Rev. B* **39(4)**: 2269-2278.

Jurelo A R, Andrade S, Jardim R F, Fonseca F C, Torikachvili M S, Lacerda A H and Ben-Dor L (2007) Effect of Ir substitution in the ferromagnetic superconductor $RuSr_2GdCu_2O_8$. *Physica C* **454(1-2)**: 30-37.

Kadin A M, Epstein K and Goldman A M (1983) Renormalization and the Kosterlitz-Thouless transition in a two-dimensional superconductor. *Phys. Rev. B* **27(11)**: 6691-6702.

Kamal S, Liang R, Hosseini A, Bonn D A and Hardy W N (1998) Magnetic penetration depth and surface resistance in ultrahigh-purity $YBa_2Cu_3O_{7-\delta}$ crystals. *Phys. Rev. B* **58(14)**: R8933-R8936.

Kanai M, Kawai T and Kawai S (1990) Superconducting superlattices: Verification of two-dimensional nature in high $T_C$ $Bi_2Sr_2(Ca_{1-x}Y_x)Cu_2O_8$ superconductors. *Appl. Phys. Lett.* **57(2)**: 198-200.

Kim D H, Goldman A M, Kang J H and Kampwirth J H (1989) Kosterlitz-Thouless transition in $Tl_2Ba_2CaCu_2O_8$ thin films. *Phys. Rev. B* **40(13)**: 8834-8839.

Kini A M, Geiser U, Wang H H, Carlson K D, Williams J M, Kwok W K, Vandervoort K G, Thompson J E, Stupka D L and Jung D (1990) A new ambient-pressure organic superconductor, κ–(ET)$_2$Cu[N(CN)$_2$]Br, with the highest transition temperature yet observed (inductive onset $T_C$ = 11.6 K, resistive onset = 12.5 K). *Inorg. Chem.* **29(14)**: 2555-2557.

Kitano H, Ohashi T, Maeda A and Tsukada I (2006) Critical microwave-conductivity across the phase diagram of superconducting $La_{2-x}Sr_xCuO_4$ thin films. *Phys. Rev. B* **73(9)**: 092405(1-4).

Kosterlitz J M and Thouless D J (1973) Ordering, metastability and phase transitions in two-dimensional systems *J. Phys. C: Solid State Phys.* **6(7)**: 1181-1203.

Lewis H W (1956) Two-fluid model of an "energy-gap" superconductor. *Phys. Rev. B* **102(6)**: 1508-1511.
17